# Monitoring and Maintenance of Telecommunication Systems: Challenges and Research Perspectives

Lakmal Silva[1] and Michael Unterkalmsteiner[2]

[1] Ericsson AB, Karlskrona, Sweden,
`ruwan.lakmal.silva@ericsson.com`
[2] Blekinge Institute of Technology, Software Engineering Research Lab, Karlskrona, Sweden
`michael.unterkalmsteiner@bth.se`

**Abstract.** In this paper, we present challenges associated with monitoring and maintaining a large telecom system at Ericsson that was developed with high degree of component reuse. The system constitutes of multiple services, composed of both legacy and modern systems that are constantly changing and need to be adapted to changing business needs. The paper is based on firsthand experience from architecting, developing and maintaining such a system, pointing out current challenges and potential avenues for future research that might contribute to address them.

## 1   Introduction

The Telecom industry has undergone a substantial transformation in the last few years, working towards fifth generation (5G) networks [1]. Ericsson is no exception and has been experiencing an increased pressure to lower operational costs and quickly respond to market changes. Therefore, some development units within Ericsson made a strategic decision in 2013 to increase the reusability of components from the existing portfolio, preserving the investment of many years of tuning and debugging [2]. With this directive, a new system was built reusing four products from the portfolio. This new system is designed to manage the life cycle of virtual resources, such as virtual machines and networks. Reuse certainly sped up the development, reduced time to market along with the added benefits of quality, reused software architecture, infrastructure, and domain knowledge [3]. However, at the same time, we encountered challenges related to the interoperability of legacy products, integrating reuse practices into the development process, and the deployment of new services into cloud environments. We provide firsthand insight into these challenges and point out avenues for future research that may address these challenges.

The remainder of the paper is structured as follows. Section 2 briefly reports on related work. Section 3 introduces the industrial context of the challenges which are further expanded in Section 4. Section 5 concludes the paper.

## 2    Related Work

Khadka et al. studied legacy to SOA evolution and concluded that reverse engineering is the most common technique for legacy system understanding [4] and wrapping is the most common implementation technique. Almonaies et al. [2] surveyed the approaches to moving legacy systems to the SOA environment with the help of redevelopment, replacement, wrapping and migration strategies. Apart from SaaS migration, cloud migration is an emerging research area with various aspects of legacy-to-cloud migration [5]. Jamshidi et al. provide an interesting comparison of SOA and cloud migration in the drivers, provisioning, design principles and crosscutting concerns perspectives. Toffetti et al. [6] propose a new architecture that enables scalable and self-managing applications in the cloud while Pahl et al. surveyed cloud container technologies and architectures [7] that offer lightweight virtualization.

Balalaie et al. described the incremental migration and architectural refactoring of a commercial mobile back-end as a service to microservices architecture with the help of DevOps [8]. Finally, Varghese and Buyya discussed new trends and research directions in next generation cloud computing [9] while Xavier and Kantarci surveyed challenges and opportunities for communication and network enablers for cloud-based services [10].

## 3    Industrial Context

The system we discuss in this paper was designed to manage the life-cycle of virtual resources, such as virtual machines and networks. The reuse of software components meant that the new system would be composed of complex subsystems, even though the percentage of reused functionality from each subsystem was low (between 20-30%). It was not possible to extract only the services that were needed, as each reused subsystem was built as a monolithic application. The functionality of the new system had to be mapped to fit into the existing architecture, rather than designing a system that best suits the business requirements. This led to an architectural degradation and consequences in the systems runtime characteristics which are explained next.

### 3.1    System Architecture

As illustrated in **Fig. 1**, the reused subsystems expose different inbound interfaces. The REST interface in VM1 is the main entry point into the new system. This interface can be used to build customer specific services without having to change the core system. The main service provided by VM1 is to slice the order into fine grained requests to be orchestrated towards VM3, and to store the necessary data in the database located in VM2. Since these subsystems have been developed independently in different timelines by different development units, the protocols and technologies used vary (REST in VM1, SQL in VM2 and SOAP in VM3). When building a solution reusing these systems, there is a significant amount of overhead due to protocol transformations between the involved services.

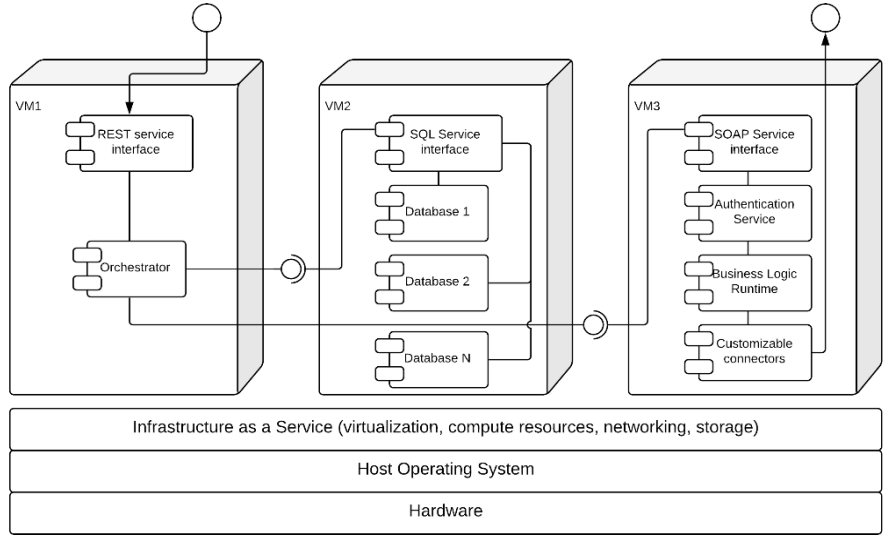

**Fig. 1.** Virtual resource manager system architecture

Similarly, the subsystems have their own ways of managing configuration. Because of this, the administrators of the system won't experience a uniform configuration management. The administrator is also expected to understand the dependencies between configurations of multiple subsystems. For instance, if a password is changed in VM1, the administrator is expected to have prior knowledge that that the password in VM2 and VM3 should also be changed. There is no centralized configuration management which distributes the configuration throughout the involved subsystems.

Different subsystems also come with a variety of databases. Their maintenance and configuration require varying techniques, forcing the administrators to acquire competency for multiple database management systems. When issues arise at customer sites, the Ericsson development unit needs to get involved, supporting the customer sites' operation and maintenance staff. This consumes time and resources that has been allocated for feature development of future releases.

### 3.2 Runtime System Characteristic

The reused subsystems were originally not developed to be run on virtualized or cloud platforms which is a strict requirement for the new system. Similar to the wrapping strategies described by Almonaies et al. [2], our approach was to wrap the legacy software into virtual machines. The resulting deployment architecture in a cloud stack is shown in **Fig. 1**. Some of the issues detailed next are the results of trying to fit legacy subsystems into modern virtualized and cloud environments.

The installation procedures of the involved subsystems vary due to their independent development by separate design units at Ericsson. These variants can be managed by using configuration and automation tools, but it is time consuming to develop a uniform framework for the new system. To add another dimension to the complexity, the subsystems are developed in different countries, driving significant delays in communication due differences in time zones.

Monitoring and recovery from process failures among the reused subsystems is not uniform. Few of the subsystems provide monitoring agents which automatically recover process failures along with sending failure and recovery notifications to the monitoring systems, whereas most subsystems provide no process recovery at all.

The complexity of the cloud stack creates problems in the areas of performance tuning and troubleshooting. Different vendors provide different layers of the cloud stack and different teams configure these layers. The performance of the systems relies on the efficiency of different layers in the cloud stack, all the way down to the hardware. The system can misbehave due to issues in any of layer of the stack, forcing support teams from each layer to get involved to resolve issues; this process can be very time consuming.

## 4      Challenges and Research Directions

Building new services/systems by reusing existing products may seem like a good approach to go to market faster and to tap into new domains with a lower cost. However, this approach comes at a cost in the long run if these systems do not evolve along technological advancements. **Table 1** identifies the key challenges, specifically when reusing large legacy systems. Next, we discuss these challenges, pointing out potential avenues for further research.

**Ch1:** Existing systems are adapted to provide services that were not intended to be supported by the original system, which results in issues being discovered very late in the development cycle and at customer sites. Another issue with the current architecture is the overhead involved in the overall system, as different subsystems are required to perform protocol transformations.

**Table 1.** Challenges related to legacy system reuse

| Id  | Challenge | Description and specifics |
|-----|-----------|---------------------------|
| Ch1 | Interoperability of legacy and new services | The reused subsystems are a mixture of both legacy systems and relatively modern systems which creates interoperability issues. |
| Ch2 | Development process for system reuse | Even though modern development processes such as DevOps have been adopted in the development phase, it is not efficient due to the characteristics of the reused subsystems. |
| Ch3 | Deployment and orchestration of services | Issues related installation, monitoring, scaling of the system in virtualized and cloud environments. |

**Ch1.1:** The system is required to comply with certain standards such as ETSI GS NFV-MANO demanded by customers. This adds to the complexity of the system architecture, as standardization was not regarded important when the legacy systems were initially designed.

*Research directions:* Previous research has illustrated strategies for migrating monolithic to microservices-based architectures that allow for flexible reuse [8,11]. The environment at Ericsson requires however adaptations to these strategies, since development is globally distributed, and the product needs to be standard compliant.

**Ch2:** The development unit at Ericsson responsible for the system has already adopted DevOps practices such as Continuous Integration and Deployment (CI/CD). However, the subsystems are so complex and large such that the lengthy CI/CD cycle renders development inefficient.

**Ch2.1:** The development teams are distributed in at least five geographically dispersed locations. To facilitate management, the teams were organized as feature teams, that is, when a feature was handed over to a design team, they were responsible for the end-to-end development of the feature, which involved adaptations to a few complex subsystems. Due to the complexity of each subsystem, there are no subsystem experts in teams, resulting in misuse of the architecture of the subsystems and therefore contributing to the degradation of the overall system architecture.

*Research directions:* DevOps practices need to be improved to facilitate the growing complexity in platforms and IT operations to prevent DevOps practices becoming an overhead in the development process. As others have observed [12], insufficient communication, company culture, legal constraints and heterogeneous environments can prevent successful DevOps practices adoption. Similar causes can be observed at Ericsson and dedicated solutions need to be investigated.

**Ch3:** The reused subsystems were not designed to be deployed into virtual environments, and the developed wrappers and frameworks soon became complex and are hard to manage.

**Ch3.1:** The demands on the operation and maintenance staff rose to the same degree as the complexity of the overall system increased. This contributes to increased system maintenance and troubleshooting time, due to inhomogeneous and scattered information in subsystems.

**Ch3.2:** When the system is deployed into cloud environments, it has also to be continuously monitored and managed [6]. The basis for efficient management of cloud-based services is to slice the monolithic applications into microservices [6]. The current strategy to wrap monolithic subsystems into a single VM makes it difficult to extract subsystems into their own services and be managed independently.

*Research directions:* There is a need to provide support for the operation and maintenance staff to interpret and analyze the data collected in production. While this data is useful for troubleshooting, it requires expertise that is not always available and costly. One possibility is to combine big data analytics and machine learning to support operation staff in decision making by filtering and identifying patterns in data [13,14].

## 5    Conclusions and Future Work

The demand for new and more features, at a faster pace and at a lower cost, has affected all high-tech industries. This requires that well established players embrace this change and identify strategies that allow them to reuse existing technologies, adapt them to new requirements, and can provide thereby new services on time, quality and cost. In this paper, we have illustrated some of the challenges related to this transition, seen through the lens of an experienced architect at Ericsson. Future work is targeted at providing a systematic, multi-perspective investigation to prioritize the challenges and guide solution development.

## References

1. Chen, S., Zhao, J.: The requirements, challenges, and technologies for 5G of terrestrial mobile telecommunication. IEEE Communications Magazine. 52, 36–43 (2014).
2. Almonaies, A.A., Cordy, J.R., Dean, T.R.: Legacy System Evolution towards Service-Oriented Architecture. In: International Workshop on SOA Migration and Evolution (SOAME). IEEE, Madrid, Spain (2010).
3. Mohagheghi, P., Conradi, R.: An Empirical Investigation of Software Reuse Benefits in a Large Telecom Product. ACM Trans. Softw. Eng. Methodol. 17, 13:1–13:31 (2008).
4. Khadka, R., Saeidi, A., Idu, A., Hage, J., Jansen, S.: Legacy to SOA Evolution: A Systematic Literature Review. In: Migrating Legacy Applications: Challenges in Service Oriented Architecture and Cloud Computing Environments. pp. 40–70. IGI Global (2013).
5. Jamshidi, P., Ahmad, A., Pahl, C.: Cloud Migration Research: A Systematic Review. IEEE Transactions on Cloud Computing. 1, 142–157 (2013).
6. Toffetti, G., Brunner, S., Blöchlinger, M., Spillner, J., Bohnert, T.M.: Self-managing cloud-native applications: Design, implementation, and experience. Future Generation Computer Systems. 72, 165–179 (2017).
7. Pahl, C., Brogi, A., Soldani, J., Jamshidi, P.: Cloud Container Technologies: a State-of-the-Art Review. IEEE Transactions on Cloud Computing. PP, 1–1 (2017).
8. Balalaie, A., Heydarnoori, A., Jamshidi, P.: Microservices Architecture Enables DevOps: Migration to a Cloud-Native Architecture. IEEE Software. 33, 42–52 (2016).
9. Varghese, B., Buyya, R.: Next generation cloud computing: New trends and research directions. Future Generation Computer Systems. 79, 849–861 (2018).
10. Xavier, G.P., Kantarci, B.: A survey on the communication and network enablers for cloud-based services: state of the art, challenges, and opportunities. Ann. Tel. 1–24 (2018).
11. Dragoni, N., Dustdar, S., Larsen, S.T., Mazzara, M.: Microservices: Migration of a Mission Critical System. (2017).
12. Riungu-Kalliosaari, L., Mäkinen, S., Lwakatare, L.E., Tiihonen, J., Männistö, T.: DevOps Adoption Benefits and Challenges in Practice: A Case Study. In: Product-Focused Software Process Improvement. pp. 590–597. Springer (2016).
13. Zaman, F., Hogan, G., Meer, S.V.D., Keeney, J., Robitzsch, S., Muntean, G. m: A recommender system architecture for predictive telecom network management. IEEE Communications Magazine. 53, 286–293 (2015).
14. Parwez, M.S., Rawat, D.B., Garuba, M.: Big Data Analytics for User-Activity Analysis and User-Anomaly Detection in Mobile Wireless Network. IEEE Transactions on Industrial Informatics. 13, 2058–2065 (2017).